\documentclass[aps,pra,twocolumn,showpacs]{revtex4-1}

\usepackage{amsmath}
\usepackage{graphicx}
\usepackage{times}
\usepackage{amssymb}
\usepackage{hyperref}
\usepackage{textcomp}
\usepackage{color}

\begin{document}
\title{Raman-assisted Rabi resonances in two-mode cavity QED}
\author{P. Gr\"unwald}
\email{Electronic address: peter.gruenwald2@uni-rostock.de}
\author{S. K. Singh}
\author{W. Vogel}
\affiliation{Arbeitsgruppe Quantenoptik, Institut f\"ur Physik, Universit\"at Rostock, D-18055 Rostock, Germany}
\date{\today}

\begin{abstract}
The dynamics of a vibronic system in a lossy two-mode cavity is studied, with the first mode being resonant to the electronic transition and the second one being nearly resonant due to Raman transitions. We derive analytical solutions for the dynamics of this system.  For a properly chosen detuning of the second mode from the exact Raman resonance, we obtain conditions that are closely related to the phenomenon of Rabi resonance as it is well known in laser physics. Such resonances can be observed in the spontaneous emission spectra, where the spectrum of the second mode in the case of weak Raman coupling 
is enhanced substantially.
\end{abstract}

\pacs{42.50.Pq, 37.30.+i, 42.50.Ct}

\maketitle

\section{Introduction}
The field of cavity QED is one of the main areas of quantum optical research nowadays~\cite{Walther06}. 
Cavities enhance the interaction time between an electromagnetic mode and an atomic system, which increases the coherence time of the atom-radiation system. This allows superior measurements of quantum correlations and their properties compared to experiments in free space. Many quantum optical phenomena in the emission of cavities could be predicted, such as photon antibunching and squeezed light~\cite{Carmichael85}, stationary  occupation inversion~\cite{Savage}, or subnatural linewidths~\cite{Carmichael89}. Cavities are also used to ``slow down'' or even freeze radiation fields via interaction on a sufficiently long time scale~\cite{Lukin}. More recently, the dynamics of single-photon wave packets in the strong-coupling regime has been theoretically studied~\cite{Difidio} and also observed in experiments~\cite{Rempe2}.
Due to the manifold of interesting features, cavity structures are one of the fundamental resources for the technical implementation of quantum information algorithms.

The strong atom-radiation coupling regime of cavity QED has been studied in various systems. In experiments, microwave and rf cavities are coupled to Rydberg atoms of large principal quantum number~\cite{Rydberg}, which propagate through the cavity.
Strong atom-field couplings have also been achieved in optical cavities~\cite{Rempe}.
Another interesting development is the combination of ion trapping and cavity QED~\cite{Lange1,Lange2}.
The fabrication of high-quality cavities and ion traps allows one to keep an interacting atom at a very precise position inside the cavity for very long times~\cite{Walther04}. This yields experimental realizations for many of the above-mentioned theoretical predictions.
More recently, based on Fabry-Perot-interferometry~\cite{Yablonovitch}, semiconductor microcavities have been developed, where excitons in quantum dots act as atomic systems~\cite{Gerard,Michler}.

In most cases, cavity QED systems describe the interaction of atoms with a single-mode cavity. However, in general, there exist more modes. A typical scenario is the quasiresonant interaction of an atomic transition with a single cavity mode, which dominates the coupling. The influence of other modes can often be neglected. Due to their off-resonance, they contribute in an oscillatory manner to the interaction, which averages to zero over sufficiently large times. 

Depending on the structure of the atomic system, it may be possible to excite 
an additional degree of freedom, such as a vibrational excitation of a molecule or a
trapped ion. In such situations, additional cavity modes may become relevant because of Raman resonances. The Raman resonances, however, essentially leave the Rabi oscillations nearly unchanged, since the interaction on the Raman resonance and on the electronic transition are in phase~\cite{Papadopoulos}. Typically the dynamics 
of the Raman-resonant cavity field follows that being resonant on the bare electronic transition. The situation changes drastically when the Raman transition is quasiresonant, such that the detuning corresponds to the Rabi frequency of the strongly coupled electronic transition. For such a scenario, will we use throughout our paper the term Raman-assisted Rabi resonance (RARR). In such a case, an irregular behavior of the two cavity modes was predicted~\cite{Dung}. Phenomena of this type may also play some role in semiconductor microcavities, where phonons may be excited.

In the present paper, we study a vibronic system in a two-mode cavity, where the first mode is resonant to the electronic transition. The second mode is Raman quasiresonant, which leads to vibrational excitations. The remaining detuning from the exact Raman resonance is a free parameter. We show that for a detuning of the order of the Rabi-oscillation frequency, the Raman-assisted mode becomes resonantly driven by the Rabi oscillation. This drastically changes the dynamics of the radiation fields in the two modes. 
Under these conditions, the excitation of the Raman-assisted mode can 
significantly exceed that of the resonant mode, even when the Raman coupling is much weaker than the pure electronic one.

The paper is organized as follows. In Sec.~\ref{sec.model} we consider the models of one- and two-mode cavities. Section~\ref{sec.RARR} deals with the solution of the dynamics of the two-mode cavity in the case of RARR. In Sec.~\ref{sec.out} we study the spectral properties of the radiation coupled out of the cavity.
A summary and some conclusions are given in Sec.~\ref{sec.SaC}.

\section{The model}\label{sec.model}
In this section, the effects of a single-mode cavity interacting with an atom are briefly examined, before we introduce the studied two-mode cavity system. For all considered cases, two conditions are fixed. First, we have no more than one optical excitation in our system, i.e., no external pumping. Second, initially the system is excited only in the electronic state, whereas the cavity modes are in the ground state.

\subsection{One-mode cavity}
Let us first consider an atom in a single-mode cavity. 
Such models have been extensively studied; see, e.g.,~\cite{CarmichaelBook2}. Hence we will only briefly recall the main results of the treatment in order to introduce the notation and to compare the results with those for the two-mode cavities to be studied in our paper.

For an excited atom in an undamped one-mode cavity which is initially in the vacuum state, the study of the time evolution requires only two quantum states: $|E\rangle=|2,0\rangle$ represents the excited atom and no excitation in the cavity mode, and $|G\rangle=|1,1\rangle$ represents the atom in the ground state and one photon in the cavity. The Hamiltonian of the system without losses may be written in the Schr\"odinger picture as
\begin{align}
	\hat H_0&=\hbar\omega_{21}(\hat A_{22}{+}\hat a^\dagger\hat a){+}\hbar\delta\omega_a\hat a^\dagger\hat a{+}\hbar g_a(\hat a^\dagger\hat A_{12}{+}\hat A_{21}\hat a),
\end{align}
where $\omega_{21}$ is the transition frequency of the atom, $\delta\omega_a=\omega_a-\omega_{21}$ is the detuning between the cavity mode
and the atom, and the atom-field coupling strength is $g_a$. 
The operators $\hat A_{kl}$ $(k,l=1,2)$ and $\hat a$ are the atomic flip operators
and the photon annihilation operator of the intracavity field, respectively.

To include the atomic and cavity losses in the Schr\"odinger picture, we apply the quantum-trajectory approach; for details see, e.g.,~\cite{Carmichael89}. Hence we arrive at the non-Hermitian Hamiltonian,
\begin{equation}
	\hat H_{0,\text{L}}=\hat H_0-i\hbar\frac{\kappa}{2}\hat a^\dagger\hat a-i\hbar\frac{\Gamma}{2}\hat A_{22},
\label{eq.losses}
\end{equation}
where $\Gamma$ is the atomic decay rate and $\kappa$ the cavity damping rate.
The state $|\psi(t)\rangle$,
\begin{equation}
	|\psi(t)\rangle=e^{-i\omega_{21}t}C_E(t)|E\rangle+e^{-i\omega_{a}t}C_G(t)|G\rangle,
\end{equation}
describes the evolution in the time interval before the photon is coupled out of the cavity. 
The initial condition is $|\psi(0)\rangle=|E\rangle$; $C_E(t)$ and $C_G(t)$ are the 
slowly varying probability amplitudes.

The solution of the Schr\"odinger equation reads
\begin{align}
	C_E(t)&=\frac{g_a}{\Omega_{\rm R}}\cos(\Omega_{\rm R}t+\phi)e^{-\frac{1}{4}(\kappa+\Gamma)t},\label{eq.singCE}\\
	C_G(t)&=-i\frac{g_a}{\Omega_{\rm R}}\sin(\Omega_{\rm R}t)e^{-\frac{1}{4}(\kappa+\Gamma)t},\\
	\Omega_{\rm R}^2&=g_a^2-\left(\frac{\kappa-\Gamma-2i\delta\omega_a}{4}\right)^2,\\
	\tan\phi&=-\frac{\kappa-\Gamma+2i\delta\omega_a}{4\Omega_{\rm R}}.\label{eq.singphi}
\end{align}
For strong coupling ($g_a\gg\delta\omega_a,\Gamma,\kappa$), we have a resonant Rabi oscillation between $|G\rangle$ and $|E\rangle$ with frequency $\Omega_{\rm R}\approx g_a$. 
The norm of $|\psi(t)\rangle$ decreases exponentially with time, which is caused by $\kappa$ and $\Gamma$. One may summarize these results roughly as follows: for perfect resonance and no losses, we have complete Rabi oscillations between the states $|E\rangle$ and $|G\rangle$. In the general case, damping and frequency shifts occur. When examining the spontaneous emission spectrum, we find a doublet of peaks in the strong-coupling regime. The width of the split lines is determined by the emission rates $\Gamma$ and $\kappa$, with the splitting being $2\Omega_{\rm R} \approx 2 g_a$.

\subsection{Two-mode cavity}

\begin{figure}[h]
\includegraphics[width=5cm]{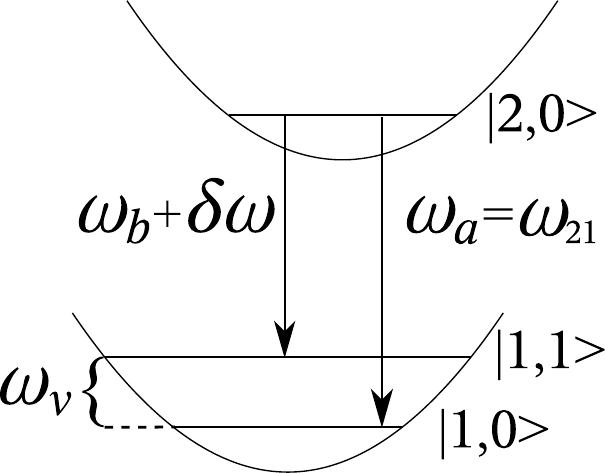}
\caption{Sketch of the vibronic system interacting with the two-mode cavity. The states $|k,l\rangle$
label the electronic and vibrational states, $k=1,2$ and $l=0,1$, respectively.
}\label{fig.System}
\end{figure}

The basic structure of the system under study is shown in Fig.~\ref{fig.System}. We consider a vibronic system in a cavity; the bare electronic transition of the former couples resonantly to one cavity mode $a$, but is off-resonant to the other modes. Via creation of a vibrational quantum, however, the corresponding vibronic transition can become nearly resonant with a second mode $b$ of lower energy.
In the low-temperature regime, vibrational excitations in the excited electronic state can be neglected.
We also may ignore the off-resonant coupling of the vibrationless transition to the second mode.

Based on these assumptions, the dynamics of our system can be described by the following three quantum states. First, in $|E\rangle=|2{,}0{,}0{,}0\rangle$, the electronic state is excited, the vibrational mode is in the ground state, and the two cavity modes are in the vacuum state. Second, $|G\rangle=|1{,}0{,}1{,}0\rangle$ describes the vibronic system in the ground state and a photon in the cavity mode $a$. Third, in $|F\rangle=|1{,}1{,}0{,}1\rangle$, the vibronic system is in the electronic ground state with a vibrational excitation and one photon in the cavity mode $b$.

In the following we will show that the most interesting situation will not be that of perfect Raman resonance with the $b$~mode, where the vibrational energy directly fills the gap between the electronic transition and that mode. For this purpose, the detuning from the Raman resonance will be included as a free parameter. 
Note that this free parameter, $\delta\omega$, is easily controlled by the cavity length.
We will find situations for which the evolutions of the excitations of the two cavity modes completely differ from the situation for exact Raman resonance, as well as from the case of one resonant and one far-off-resonant mode. 
In particular, the coupling of the vibronic system with the mode $b$ will substantially increase in a RARR scenario. 
In this situation, the Raman resonance is replaced by a resonance condition which includes the Rabi frequency describing the interaction of the vibronic system with the $a$~mode.

For the system without losses the Hamiltonian reads
\begin{align}
	\hat{H}=&\hbar\omega_{21}(\hat A_{22}+\hat a^\dagger\hat a+\hat b^\dagger\hat v^\dagger\hat b \,\hat v)-\hbar\delta\omega\hat b^\dagger\hat v^\dagger\hat b\hat v\nonumber\\
	&+\hbar \hat A_{21}\left(g_a\hat a+g_b\hat b\hat v\right)+\text{h.c.},\label{eq.Hamiltonian}
\end{align}
where $\delta\omega=\omega_{21}-\omega_b-\omega_\nu$ is 
the detuning from the Raman resonance, with the mode frequency $\omega_b$ of the $b$ mode, and the vibrational frequency $\omega_\nu$. For the structure of the vibration-assisted coupling to the $b$ mode we refer 
to~\cite{DiFidio,Knight}.
The coupling strength $g_b$ describes the atom-field coupling on the vibrational sideband. The operators $\hat b$ and $\hat v$ represent the annihilation operators of the $b$ mode and of a vibrational excitation, respectively.
The actual value for $g_b$ can be deduced from the precise structure of the system under study. For our purpose, we will assume that $g_b\ll g_a$, which holds true for systems with a weak vibronic coupling.
The state $|\psi(t)\rangle$ can be written as
\begin{equation}
 \begin{split}
	|\psi(t)\rangle=&e^{-i\omega_{21}t}C_E(t)|E\rangle+e^{-i\omega_{a}t}C_G(t)|G\rangle\\
	&+e^{-i(\omega_{21}-\delta\omega)t}C_F(t)|F\rangle,
 \end{split}
\end{equation}
where $C_K(t)$ ($K=E,G,F$) are the occupation probability amplitudes of the states as defined above.

For perfect Raman resonance ($\delta\omega=0$), the Hamiltonian~(\ref{eq.Hamiltonian}) reduces to
\begin{equation}
\begin{split}	
\hat{H}=&\hbar\omega_{21}(\hat A_{22}+\hat a^\dagger\hat a+\hat b^\dagger\hat v^\dagger\hat b\hat v)\\
	&+\hbar \hat A_{21}\left(g_a\hat a+g_b\hat b\hat v\right)+\text{h.c.}.\label{eq.RamanHamiltonian}
\end{split}
\end{equation}
In this case, the solutions for the coefficients are readily obtained as
\begin{align}
	C_E(t)&=\cos(\Omega_\text Rt),\\
	C_G(t)&=-i\frac{g_a}{\Omega_\text R}\sin(\Omega_\text Rt),\\
	C_F(t)&=-i\frac{g_b}{\Omega_\text R}\sin(\Omega_\text Rt),\\
	\Omega_\text R^2&=g_a^2+g_b^2.
\end{align}
This corresponds to a three-level system, with an effective Rabi frequency determined by the couplings of the two vibronic transitions with the two cavity modes. In this case, the occupation probabilities of the two modes obey exactly the same dynamics. They are only weighted by the different (squared) coupling strengths of the corresponding cavity modes to the different vibronic transitions,
\begin{equation}
	\frac{|\langle F|\psi(t)\rangle|^2}{|\langle G|\psi(t)\rangle|^2}=\frac{|C_F(t)|^2}{|C_G(t)|^2}=\frac{g_b^2}{g_a^2}.\label{eq.pureRaman}
\end{equation}
For a weak vibronic coupling, the occupation of the $b$~mode is thus very small compared to that of the $a$~mode. Note that in a somewhat different context, this solution has been used for cavity systems with degenerate cavity modes~\cite{Papadopoulos}.

\section{Raman-assisted Rabi Resonances}\label{sec.RARR}
As already stated above, we are interested in the effect of a Raman-assisted coupling of the atomic system to the $b$~mode, under conditions when the vacuum Rabi splitting due to the coupling with the $a$~mode is relevant, leading to so-called RARR. To our best knowledge, such scenarios have not been considered yet in the context of cavity QED.
In this section, we will study the dynamics without losses. We will visualize the resulting dynamics and explain the physics behind RARR.

The mathematical structure of a lossless two-mode cavity interacting with an atom has been studied in~\cite{Dung}. However, the authors did not consider a physical system realizing the studied behavior. Even more importantly, they only studied some special conditions. Hence the authors could not provide a detailed interpretation of the physics and they did not consider the RARR, in which we are interested here. As we will show below, in our system, an interpretation of the dynamics is straightforward, both for our conditions of RARR and for those considered in~\cite{Dung}.

Recalling Eq.~(\ref{eq.Hamiltonian}), the Schr\"odinger equation for this system leads to a third-order differential equation for the coefficients of the state $|\psi(t)\rangle$, such as
\begin{equation}
	\left[\frac{d^3}{dt^3}{-}i\delta\omega\frac{d^2}{dt^2}{+}(g_a^2{+}g_b^2)\frac{d}{dt}{-}ig_a^2\delta\omega \right]C_E(t){=}0.\label{eq.diff3}
\end{equation}
These equations of motion can be easily solved in general, but the analytical expressions do not yield much insight into the physical phenomena. Thus we focus on some more direct descriptions of the system at hand. An exponential ansatz $e^{\lambda_i t}$ for the solution leads to three purely imaginary solutions, as we included no losses. They represent the three, generally incommensurate, eigenfrequencies of the system. 

\begin{figure}[h]
\centering
\includegraphics[width=8cm]{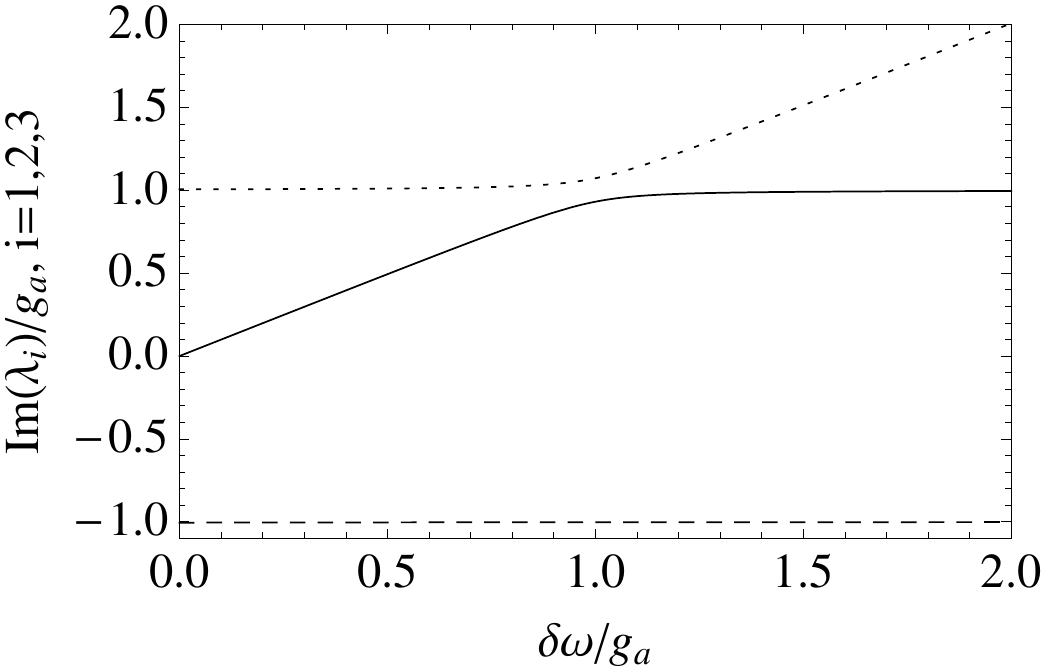}
\caption{Imaginary parts of the solutions $\lambda_1$ (dashed line), $\lambda_2$ (solid line), and $\lambda_3$ (dotted line) as functions of $\delta\omega$. They are obtained from an exponential ansatz for Eq.~(\ref{eq.diff3}) for the parameter $g_a/g_b=10$.} 
\label{fig.freq3}
\end{figure}

In Fig.~\ref{fig.freq3} they are shown as functions of $\delta\omega$ for $g_a/g_b=10$.
For convenience we label them as $\lambda_i$ ($i=1,2,3$) for further discussions.
One can see that for almost no detuning, we reobtain the results for a perfect Raman resonance; see Eqs.~(\ref{eq.RamanHamiltonian})--(\ref{eq.pureRaman}). 
For very large detuning, $\delta\omega\gg g_a$, the $b$ mode is far off-resonant, even from the vibrational state of the atom. Hence it is suppressed, as in the usual case of single-mode cavity QED. The Rabi frequency of the oscillation between the electronic state and the $a$ mode becomes $g_a$, independently of $g_b$.

Around $\delta\omega\approx g_a$, we observe in Fig.~\ref{fig.freq3} an avoided crossing between $\lambda_2$ and $\lambda_3$. 
In the region of this crossing we have $\lambda_2\approx\lambda_3\approx-\lambda_1\approx ig_a$. Hence the main Rabi cycle between the atom and the $a$ mode is not suppressed. Parallel to this transition, the close frequency branches $\lambda_2$ and $\lambda_3$ lead to an independent oscillation with the frequency $\Im(\lambda_3-\lambda_2)/2$, with $\Im$ being the imaginary part. As this frequency is also not suppressed, we obtain two independent oscillations in different frequency ranges: the fast Rabi oscillation with approximately $g_a$ and the much slower one with the frequency $\Im(\lambda_3-\lambda_2)/2$.

The three occupation probability amplitudes, given as a sum of the three resulting exponentials with corresponding prefactors, are of the form
\begin{align}
	C_E(t)&=\sum_{n=1}^3\frac{(\lambda_n-i\delta\omega)\lambda_n}{g_a^2+g_b^2+(3\lambda_n-2i\delta\omega)\lambda_n}e^{\lambda_nt},\label{eq.C_E}\\
	C_F(t)&=\sum_{n=1}^3\frac{-ig_b\lambda_n}{g_a^2+g_b^2+(3\lambda_n-2i\delta\omega)\lambda_n}e^{\lambda_nt},\label{eq.C_F}\\
	C_G(t)&=\sum_{n=1}^3\frac{-ig_a(\lambda_n-i\delta\omega)}{g_a^2+g_b^2+(3\lambda_n-2i\delta\omega)\lambda_n}e^{\lambda_nt}.\label{eq.C_G}
\end{align}
In the region around $\delta\omega\approx g_a$, we see that the coefficients proportional to $\lambda_n-i\delta\omega$ decrease significantly. Hence the coefficient $C_F(t)$ plays an increasing  role. The main transition is expected to be the strongly coupled electronic one between $|E\rangle$ and $|G\rangle$. However, on a longer time scale, the occupation of the $b$ mode, being related to the state $|F\rangle$, may even exceed the occupation of the strongly coupled $a$ mode, due to the effect of RARR.

\begin{figure}[h]
\centering
 \includegraphics[width=8cm]{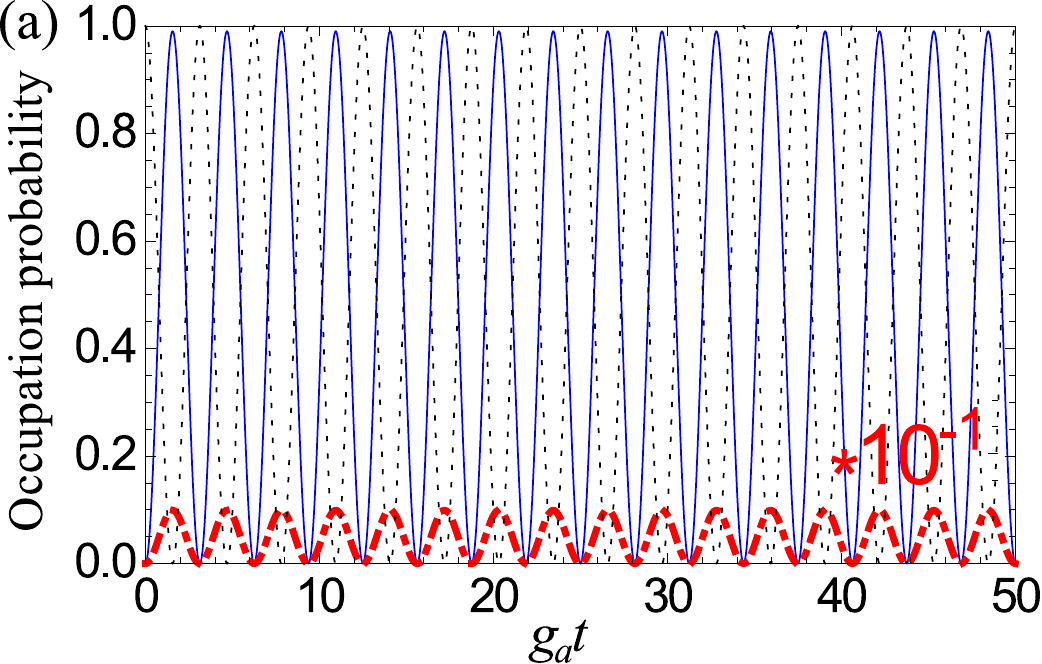}
 \includegraphics[width=8cm]{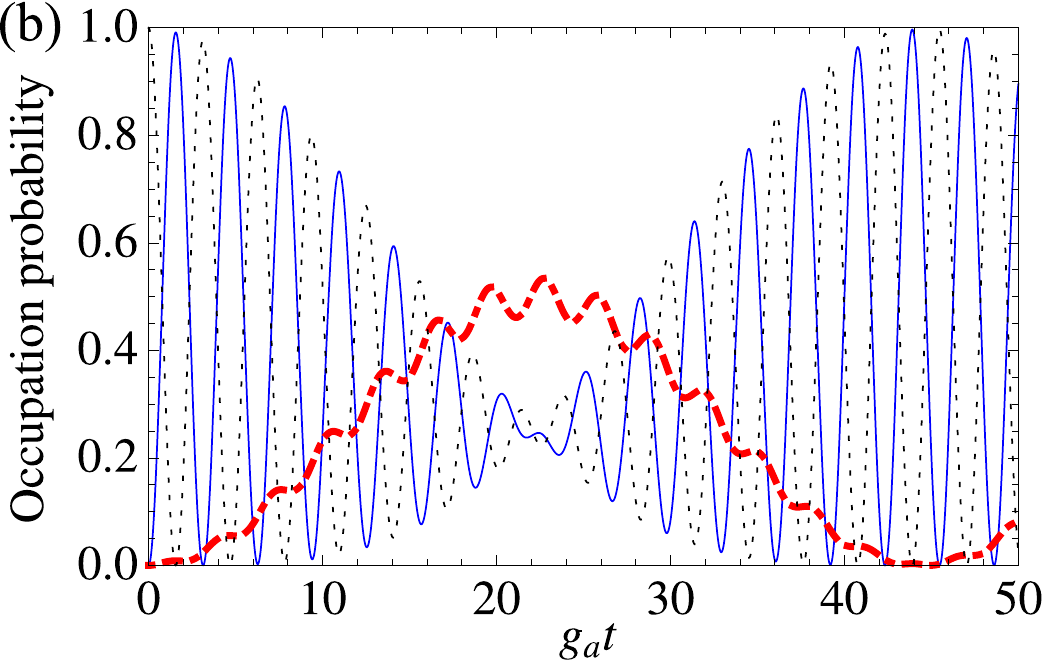}
\caption{(Color online) Occupation probabilities as a function of the scaled time $g_at$ for the atomic excited state (black dotted line), the resonant $a$ mode (blue solid line), and the $b$ mode (thick red dashed-dotted line) for $g_a/g_b=10$.  (a) The pure Raman resonance, $\delta\omega=0$, where the $b$ mode is magnified by a factor of 10 to be clearly visible. (b) The RARR scenario, with the Raman-assisted transition being detuned by $\delta\omega=g_a$.}\label{fig.Occpop1}
\end{figure}

In Fig.~\ref{fig.Occpop1} we compare the occupation probabilities $|C_K(t)|^2$ of the states $|K\rangle$ ($K=E,F,G$) for $g_a/g_b=10$, for both the perfect Raman resonance [$\delta\omega=0$, Fig.~\ref{fig.Occpop1}(a)] and for RARR [$\delta\omega=g_a$, Fig.~\ref{fig.Occpop1}(b)]. The time evolutions can be seen to differ substantially in both cases. As expected from Eq.~(\ref{eq.pureRaman}), for Raman-resonance the occupation $|C_F(t)|^2$ of the $b$ mode is shown to follow that of the $a$ mode, $|C_G(t)|^2$, but the former is smaller by two orders of magnitude. 
For the RARR scenario, all three occupation probabilities still show the typical Rabi oscillations, but for specific time intervals, the occupation of the $b$ mode even exceeds that of the $a$ mode.

The physical explanation of this situation is rather simple. The $b$ mode is quasiresonant and, in slowly varying variables, its occupation probability is oscillating with a frequency offset of  $\delta\omega$ relative to the main Rabi cycle. Thus it becomes very small in the case of Raman resonance. By choosing $\delta\omega \approx g_a$, the vibronic transition becomes resonantly driven, which is caused by the Rabi oscillation between the states $|E\rangle$ and $|G\rangle$.

This effect is well known in laser physics for two-mode laser beams. The  so-called Rabi resonances lead to an enhanced atomic excitation if the first laser mode is resonant with the atomic transition and the second one is detuned by the Rabi frequency characterizing the atom-field coupling with the first mode~\cite{Laser,LaserRabi}. However, in cavity QED such an effect, to our best knowledge, has not been considered so far. The essential difference in our case, compared to laser physics, is that we have a limited amount of energy, as there is only one optical excitation. Consequently, as the occupation of the $b$ mode increases, the occupations of the other two states are reduced, until the former becomes dominant and the Rabi oscillations nearly die out. In this case, both the atom and the $a$ mode approach the values of $|C_E(t)|^2=|C_G(t)|^2=1/4$. Over time, the process is inverted, with the $b$ mode driving the Rabi cycle between $|E\rangle$ and $|G\rangle$, and the evolution starts over again.

From this interpretation, it becomes clear why the authors of~\cite{Dung} could not explain their results. As the eigenfrequencies are incommensurate, the two oscillations do not map onto each other with rational values. This yields a  phase shift, leading to a more or less irregular evolution.
Furthermore, the authors considered the situation for $g_a/g_b=2$. In this case, the change of the occupation of the $b$ mode is large, even within one Rabi cycle. Thus
the $b$-mode occupation probability reaches its maximal value and decreases again within a few Rabi cycles. Due to the irrational relations, this maximum may appear at some random phase in the oscillation. Hence the dynamical pattern in such a case may seem to be ``chaotic like'', cf.~\cite{Dung}. For $g_b\ll g_a$, however, the change of the occupation of the $b$ mode is rather small within one Rabi cycle and the dynamics of the two modes appears to be regular.

\section{Spectral properties of the external field}
\label{sec.out}
Let us now include the losses into this system as in Eq.~(\ref{eq.losses}). The Hamiltonian then reads
\begin{align}
	\hat{H}_L=&\hat H-i\hbar\frac{\Gamma}{2}\hat A_{22}-i\hbar\frac{\kappa}{2}\hat a^\dagger\hat a-i\hbar\frac{\kappa}{2}\hat b^\dagger\hat b,
\end{align}
where $\hat H$ is the lossless Hamiltonian in Eq.~(\ref{eq.Hamiltonian}). The $b$ mode has the same emission rate $\kappa$ as the $a$ mode, as this rate is just a geometric parameter of the cavity. Note that the losses here include the out-coupling of the photon from the $b$ mode, but no decay of the vibrational quantum to be excited in the Raman-assisted transition. These excitations have usually much longer lifetimes than the electronic excited state and the intracavity photons. Thus we may disregard the decay of vibrational excitations.

Similar to the results of the single-mode calculations see Eqs.~[(\ref{eq.singCE})--(\ref{eq.singphi})], now there appear shifts in all eigenfrequencies $\lambda_i$. In the strong-coupling regime, these shifts become negligibly small. A real damping part occurs in all frequencies, with a damping rate of $(\Gamma+\kappa)/4$. In the following, we will study the photon emission properties of our system, which can be easily measured in the field outside the cavity.

There are three different decaying channels through which the photons may leave the cavity. First, it can be directly emitted from the atom out of the side of the cavity. This process occurs due to the atomic  decay with the rate $\Gamma$. In an experiment, where the experimenter usually excites the atom and is interested in the photons emitted along the cavity axis, these events represent unwanted photon losses. Second, a photon can be emitted out of the cavity mode $a$ to record a count at a detector around the frequency $\omega_a$ with rate $\kappa$. Third, it can be emitted from the $b$ mode, leading to a recorded event around the frequency $\omega_b$. This process also occurs with a rate $\kappa$. We remind the reader that the two mode frequencies $\omega_a$ and $\omega_b$ are supposed to be significantly different and thus they can be analyzed independently.

\begin{figure}[h]
\centering
 \includegraphics[width=8cm]{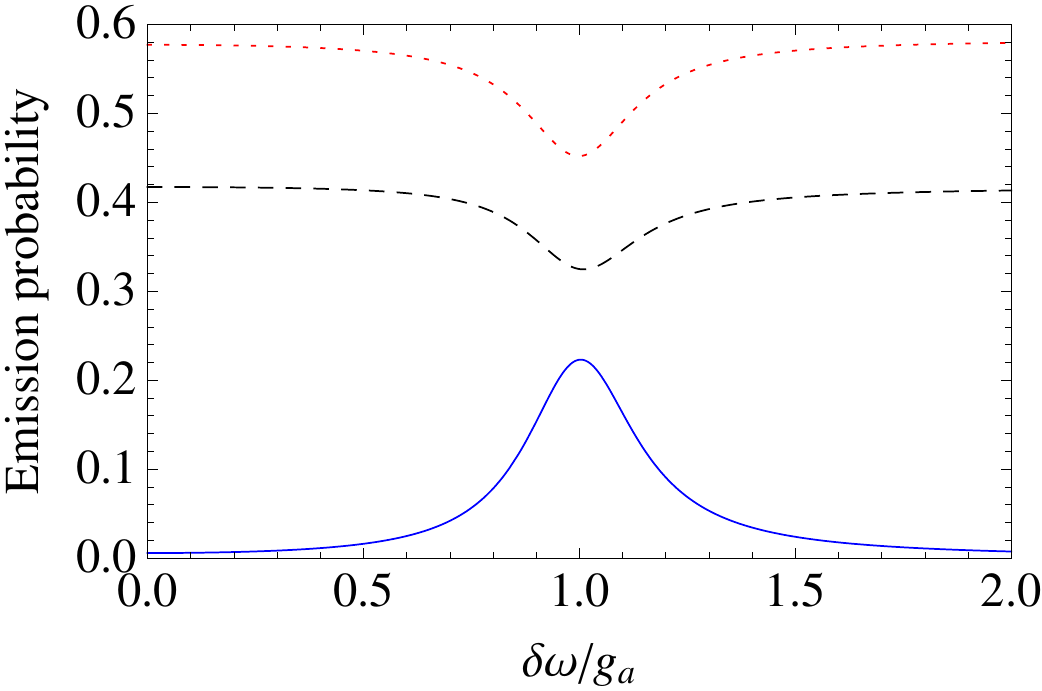}
	\caption{(Color online) The time-integrated probabilities of finding the emitted photon in the  $a$ mode (red dotted curve) or the $b$ mode (blue solid curve) are given as a function of $\delta\omega$. The dashed black curve describes the photon losses of out the side of the cavity. The parameters are chosen as $\Gamma/g_a=0.05$, $\kappa/g_a=0.07$, and $g_b/g_a=0.1$.}\label{fig.Emitprob}
\end{figure}

According to the quantum-trajectory method~\cite{Difidio}, the probabilities $p_i(t)$ of emitting the photon at time $t$ through  one of the three channels, $i=1,2,3$, are given by
\begin{align}
	p_1(t)&=\Gamma\int_0^tdt'|C_E(t')|^2,\\
	p_2(t)&=\kappa\int_0^tdt'|C_G(t')|^2,\\
	p_3(t)&=\kappa\int_0^tdt'|C_F(t')|^2.
\end{align}
Here we number the decay channels in the order of their explanation above.
The limit $t\rightarrow\infty$ yields the total emission probabilities through each decay channel. These probabilities are shown as functions of $\delta\omega$ in Fig.~\ref{fig.Emitprob}. The $a$ mode has a significantly higher emission probability than the $b$ mode, since it is on average more strongly occupied. For $\delta\omega$ close to zero, the corresponding ratio reflects that of the coupling strengths, $g_a$ to $g_b$. For large detuning, practically no emission is given from the $b$ mode, as it is far off-resonance. In the case of RARR, $\delta \omega =g_a$, the emission from the $b$ mode is enhanced by a factor of about 30 compared with exact Raman resonance. Thus, it is approximately half 
of the total emission probability of the $a$ mode.

\begin{figure}[h]
\centering
	\includegraphics[width=8cm]{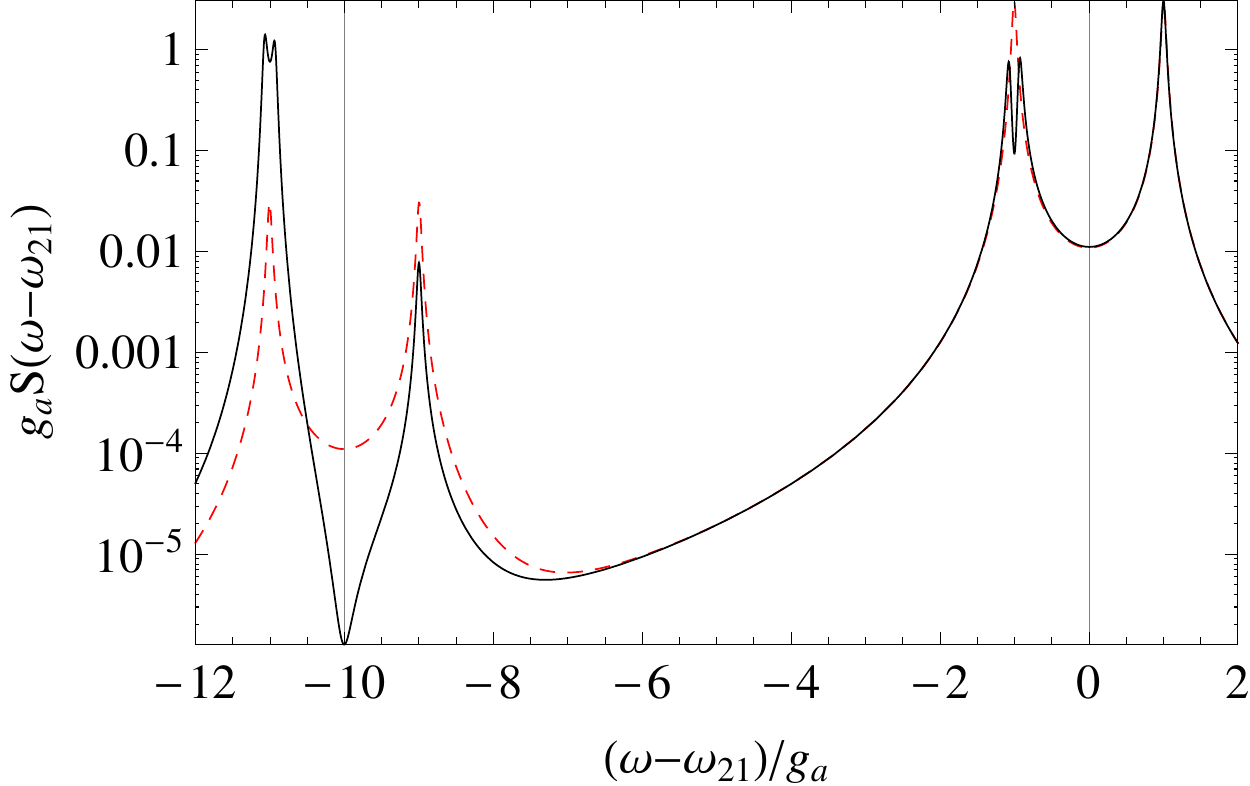}
	\caption{(Color online) Logarithmic plot of the spontaneous emission spectrum along the cavity axis for the two-mode cavity with perfect Raman resonance (red dashed line, $\delta\omega=0$) and RARR (black solid line, $\delta\omega=g_a$). The parameters are the same as in Fig.~\ref{fig.Emitprob}. The two vertical lines indicate the resonance frequencies $\omega_a$ (right) and $\omega_b$ (left) of the cavity modes.}
\label{fig.tmspec}
\end{figure}

Now we study the spontaneous emission spectrum of the described system. We calculate the time-integrated spectra as used in~\cite{Carmichael89}. In the Schr\"odinger picture, the spectrum of the two-mode cavity can be easily obtained via
\begin{align}
	S(\omega)&=\kappa/(2\pi)[S_a(\omega)+S_b(\omega)],\\
	S_a(\omega)&=\int_0^\infty \hspace{-0.3cm}dt\int_0^\infty \hspace{-0.3cm}dt'C_G^*(t)C_G(t')e^{-i(\omega-\omega_a)(t-t')},\\
	S_b(\omega)&=\int_0^\infty \hspace{-0.3cm}dt\int_0^\infty \hspace{-0.3cm}dt'C_F^*(t)C_F(t')e^{-i(\omega-\omega_b)(t-t')}.
\end{align}
The respective single-mode spectra $S_a(\omega)$ and $S_b(\omega)$ are located around their resonance frequencies. Thus we obtain well-separated spectra, provided that $\omega_a-\omega_b\gg g_a$ is fulfilled. We may again distinguish the two cases of exact Raman resonance, $\delta\omega=0$, and of RARR, $\delta\omega=g_a$.

The spectra for both scenarios are shown in Fig.~\ref{fig.tmspec}. For the exact Raman resonance, we obtain the expected frequency spectrum with two Rabi-split peaks around each mode frequency. For the RARR, however, we get a triplet structure around both mode frequencies. Both the spectra of the $a$ mode and the $b$ mode undergo a slight splitting at the low-frequency side, which is of the order of
$g_b$. As an important effect of RARR, we observe a substantial enhancement of the emission spectrum of the $b$ mode at the low-frequency side.

\section{Summary and Conclusion}\label{sec.SaC}
In summary, we have studied a two-mode cavity system with a single vibronic system inside it. One of the cavity modes, the $a$ mode, was assumed to be resonant to the bare electronic transition. The other one, the $b$ mode, is far off\-resonant with respect to the electronic transition, so that it would be ignored in standard scenarios of cavity QED. However, in our approach, the $b$ mode couples nearly resonant to a vibronic transition of the atomic system, with some detuning that can be properly adjusted. For exact Raman resonance of this mode with the corresponding vibronic transition, we reobtain the known results that the dynamics of both modes undergoes the same Rabi oscillation. For rather weak vibronic coupling, the occupation of the $b$ mode is almost suppressed compared with that of the $a$ mode. Hence for exact Raman resonance, the contribution of the former to the dynamics of the system is very small.

The effects of the $b$ mode become more important if we choose a detuning from the exact Raman resonance by the Rabi frequency for the strong interaction of the $a$ mode with the electronic transition. In this case, we obtain a Raman-assisted Rabi resonance, which substantially increases the influence of the $b$ mode on the system dynamics. The occupation of the $b$ mode can even exceed that of the strongly coupled $a$ mode, thereby draining the population of the main Rabi cycle. Over time, the occupation of the $a$ mode increases again, and so forth.

We have also studied the spontaneous emission spectrum of the output field from the two-mode cavity. Due to the Raman-assisted Rabi resonance, the spectrum around the frequency of the $b$ mode is strongly enhanced at its low-frequency side. The spectra of both cavity modes are split into triplets, with the dominant splitting being caused by the Rabi frequency driving the electronic transition through the $a$ mode. In addition, the spectra on the low-frequency sides of the two modes show another splitting with the Rabi frequency driving the interaction with the $b$ mode.
Altogether, our results clearly show that Rabi-resonance effects may strongly modify the dynamics and the emission spectra of a two-mode cavity interacting with a vibronic system.

This work was supported by the Deutsche Forschungsgemeinschaft through Grant No. SFB 652.

\end{document}